\documentclass[12pt]{article}
\usepackage{amstex}
\begin{document}
\begin{center}
\LARGE
\textbf{No contradictions between Bohmian}\\
\textbf{and quantum mechanics}\\[1cm]
\large
\textbf{Louis Marchildon}\\[0.5cm]
\normalsize
D\'{e}partement de physique,
Universit\'{e} du Qu\'{e}bec,\\
Trois-Rivi\`{e}res, Qc.\ Canada G9A 5H7\\
email: marchild$\hspace{0.3em}a\hspace{-0.8em}
\bigcirc$uqtr.uquebec.ca\\
\end{center}
\medskip
\begin{abstract}
Two recent claims by A.~Neumaier (quant-ph/0001011) and
P. Ghose (quant-ph/0001024) that Bohmian mechanics
is incompatible with quantum mechanics for 
correlations involving time are shown to be unfounded.
\end{abstract}
\section{Introduction}
Bohmian mechanics~\cite{bohm,holland} was proposed
to give a realistic and deterministic interpretation
of quantum mechanics.  It assumes that every particle
has, at any time $t$, a well-defined position on a
trajectory that obeys Newton-type equations of motion.
That motion is, however, influenced by a ``quantum
potential'' which is related to the Schr\"{o}dinger
wave function of the particle.  Trajectory parameters,
although well-defined, cannot be known exactly.  One
can only know the probability density that the particle
is at a given point, at a given time.  That probability is
equal to the absolute square of the normalized wave
function.  In this way Bohmian mechanics reproduces
exactly the statistical predictions of quantum mechanics
as regards expectation values of observables.

In recent papers, A.~Neumaier~\cite{neumaier} and
P.~Ghose~\cite{ghosea,ghoseb} have argued that although
Bohmian mechanics may well exactly reproduce the
single-time statistical predictions of quantum mechanics,
the two approaches disagree on some
multiple-time observables.  It is the purpose of this
note to show that these claims are unfounded.
\section{Neumaier's argument}
Neumaier~\cite{neumaier} considers a
one-dimensional harmonic oscillator
of angular frequency $\omega = 2\pi / T$,
in its ground state $\psi _0$.  He then 
evaluates the following expectation value, 
in Bohmian and in quantum mechanics:
\begin{equation}
\langle \psi _0 |X(t_1+T/2) X(t_1)| \psi _0 \rangle .
\label{eq2a}
\end{equation}

In quantum mechanics, $X(t_1)$ is interpreted
as the position operator in the Heisenberg
picture.  Since the position and momentum
operators in the Heisenberg picture satisfy
the same equations of motion as the classical position
and momentum, one easily shows that $X(t_1 + T/2)
= -X(t_1)$, so that
\begin{equation}
\langle \psi _0 |X(t_1 + T/2) X(t_1)| \psi _0 
\rangle _{Q} = - \langle \psi _0 |X(t_1) ^2|
\psi _0 \rangle _{Q} .
\label{eq2b}
\end{equation}
In Bohmian mechanics, on the other hand, the
momentum associated with a real wave function
(such as $\psi _0$) always vanishes.  The true
value of position is constant, which means that
$x(t_1 + T/2) = x(t_1)$.  Hence
\begin{align}
\langle \psi _0 |X(t_1 + T/2) X(t_1)| \psi _0 
\rangle _{B} &= \int dx \, |\psi _0 (x)|^2
x(t_1 + T/2) x(t_1) \notag \\
&= \int dx \, |\psi _0 (x)|^2 x(t_1)^2 \notag \\
&= \langle \psi _0 |X(t_1) ^2|
\psi _0 \rangle _{Q}.
\label{eq2c}
\end{align}
Since the right-hand side of (\ref{eq2c}) does not
vanish, Eqs.~(\ref{eq2b}) and~(\ref{eq2c})
appear incompatible.

But are they?  For this to be so, the
expectation value (\ref{eq2a}) must have the same
meaning in Bohmian as in quantum mechanics.
This remark is anticipated in 
Neumaier~\cite{neumaier}.  However, spelling it
out carefully shows, in our opinion,
that his answer is inadequate.

In Bohmian mechanics, the meaning of (\ref{eq2a}) is
straightforward.  It is the average, over the
ground state statistical ensemble, of the product of
true values of the particle position at $t_1$ and
$t_1 + T/2$.

In Copenhagen quantum mechanics, the situation is more
complicated since there is no such thing as the
true value of position (except in the special
case where the wave function is an eigenstate
of the position operator).  The meaning of (\ref{eq2a})
is related to the probability of measurement
results.  To see this, let the quantum
state at $t=0$ be represented by a density
operator $\rho$.  Suppose that position
measurements are made at $t_1$ and $t_1 + \tau$.
The probability of obtaining the results $x$
at $t_1$ and $x'$ at $t_1 + \tau$ is given 
by~\cite{wigner}:
\begin{equation}
P(x,t_1;x',t_1 + \tau) = \mbox{Tr} \left\{
|x' \rangle \langle x' | U(\tau)
|x \rangle \langle x | U(t_1) \rho U^{\dagger} (t_1)
|x \rangle \langle x | U^{\dagger} (\tau) \right\} .
\label{eq2d}
\end{equation}
Here $U$ is the time-evolution operator.  For
simplicity, we have taken position to
be discrete.

The average of the product of $X$ (the position 
operator in the Schr\"{o}dinger picture) at 
$t_1 + \tau$ and $X$ at $t_1$ is given by
\begin{align}
& \sum _x \sum _{x'} x x' P(x,t_1;x',t_1 + \tau) \notag \\ 
& \qquad = \sum _x x \, \mbox{Tr} \left\{ \left[ 
\sum _{x'} x' |x' \rangle \langle x' | \right] U(\tau)
|x \rangle \langle x | U(t_1) \rho U^{\dagger} (t_1)
|x \rangle \langle x | U^{\dagger} (\tau) \right\}
\notag \\
& \qquad = \sum _x x \, \mbox{Tr} \left\{ 
U^{\dagger} (\tau) X U(\tau)
|x \rangle \langle x | U(t_1) \rho U^{\dagger} (t_1)
|x \rangle \langle x | \right\}
\notag \\
& \qquad = \sum _x x \langle x |
U^{\dagger} (\tau) X U(\tau) |x \rangle 
\langle x | U(t_1) \rho U^{\dagger} (t_1)
|x \rangle .
\label{eq2e}
\end{align}
Let $\rho = |\psi _0 \rangle \langle \psi _0 |$
and let $\tau = T/2$, so that $U^{\dagger} (\tau)
X U(\tau) = - X$.  Then
\begin{align}
\sum _x \sum _{x'} x x' P(x,t_1;x',t_1 + \tau) &=
\sum _x x \langle x | {-X} | x \rangle
\langle x | U(t_1) | \psi _0 \rangle
\langle \psi _0 |U^{\dagger} (t_1)|x \rangle \notag \\ 
&= \sum _x (- x^2) \langle \psi _0 
|U^{\dagger} (t_1) |x \rangle
\langle x |U(t_1)| \psi _0 \rangle \notag \\
&= - \langle \psi _0 | U^{\dagger} (t_1) X^2 
U(t_1)| \psi _0 \rangle \notag \\
&= - \langle \psi _0 | X(t_1)^2 | \psi _0 \rangle .
\label{eq2f}
\end{align}
This establishes the meaning of (\ref{eq2b})
in terms of measurement results.

So we see that the expectation value
(\ref{eq2a}) has a different
meaning in Bohmian and in quantum mechanics.
The average of true values cannot be calculated
in Copenhagen quantum
mechanics, and thus cannot be compared with the
Bohmian result.  On the other hand, probabilities
of measurement results can be obtained in Bohmian
mechanics.  Successive measurements of $X$ at $t_1$
and $t_1 + \tau$, on a harmonic oscillator, can
be represented as follows.  Let $|\alpha _0 \rangle$
and $|\beta _0 \rangle$ be the initial states of two
apparatus designed for position measurement.  The
initial state of the global system made of the
harmonic oscillator and the two apparatus is
given by
\begin{equation}
\Phi (t=0) = |\psi _0 \rangle 
|\alpha _0 \rangle |\beta _0 \rangle 
= \sum _x \psi _0 (x) |x \rangle 
|\alpha _0 \rangle |\beta _0 \rangle .
\label{eq2g}
\end{equation}
Between $t=0$ and $t=t_1$, the system evolves
by an uninteresting phase factor.  Indeed the
three subsystems evolve independently, each being
in an eigenstate of its own hamiltonian.  At $t=t_1$,
the first measurement is made (we take it to be
instantaneous), so that the state is then ($t_1^+$
means immediately after $t_1$)
\begin{equation}
\Phi (t = t_1^+) = \sum _x \psi _0 (x) |x \rangle 
|\alpha _x \rangle |\beta _0 \rangle ,
\label{eq2h}
\end{equation}
where $|\alpha _x \rangle$ is a pointer state
corresponding to the value $x$ of position.
Between $t_1$ and $t_1 + \tau$, the subsystems
are decoupled, and the oscillator evolves 
following its own hamiltonian.  Therefore,
\begin{align}
\Phi [t= (t_1 + \tau)^-] &= \sum _x \psi _0 (x) 
[U(\tau) |x \rangle] 
|\alpha _x \rangle |\beta _0 \rangle \notag \\ 
&= \sum _x \psi _0 (x) \sum _{x'} 
[\langle x' |U(\tau) |x \rangle] |x' \rangle 
|\alpha _x \rangle |\beta _0 \rangle .
\label{eq2i}
\end{align}
At $t=t_1 + \tau$, the second measurement is
made and the state becomes
\begin{equation}
\Phi [t = (t_1 + \tau)^+] 
= \sum _x \psi _0 (x) \sum _{x'} 
[\langle x' |U(\tau) |x \rangle] |x' \rangle 
|\alpha _x \rangle |\beta _{x'} \rangle .
\label{eq2j}
\end{equation}
The probability that the first measurement
yields the value $x$ and the second one yields
$x'$ is equal to the probability that the first
pointer is truly at $\alpha _x$ and the second
one truly at $\beta _{x'}$.  It is given by
\begin{equation}
|\psi _0 (x) \langle x' |U(\tau)| x \rangle |^2 .
\label{eq2k}
\end{equation}
This coincides with the right-hand
side of Eq.~(\ref{eq2d}).
The alleged contradiction between (\ref{eq2b})
and (\ref{eq2c}) therefore disappears.
\section{Ghose's argument}
Ghose~\cite{ghosea,ghoseb} considers 
an experiment in which a pair
of identical particles are simultaneously
diffracted by two slits, as shown in Fig.~1.  The
particles may then be detected on a screen.
We have $L \gg a$.  We shall examine the form
of the wave function near the
screen, which is crucial to Ghose's argument.

\begin{figure}[ht]
\begin{center}
\begin{picture}(260,100)(0,0)
\multiput(30,5)(110,0){2}{\line(1,0){90}}
\multiput(17,50)(0,30){2}{\line(1,0){6}}
\multiput(20,50)(0,20){2}{\line(0,1){10}}
\multiput(230,5)(0,10){10}{\line(2,-1){10}}
\thicklines
\put(30,0){\line(0,1){15}}
\put(30,25){\line(0,1){50}}
\put(30,85){\vector(0,1){15}}
\put(230,0){\line(0,1){100}}
\thinlines
\qbezier(30,80)(120,72.5)(210,65)
\qbezier(30,50)(120,57.5)(210,65)
\qbezier(30,20)(120,42.5)(210,65)
\put(245,50){\vector(1,0){15}}
\put(210,65){\makebox(0,0){$\bullet$}}
\put(130,5){\makebox(0,0){$L$}}
\put(20,65){\makebox(0,0){$a$}}
\put(20,95){\makebox(0,0){$x$}}
\put(254,42){\makebox(0,0){$y$}}
\put(115,31){\makebox(0,0){$\vec{r} _B$}}
\put(105,64){\makebox(0,0){$\vec{r}$}}
\put(120,82){\makebox(0,0){$\vec{r} _A$}}
\end{picture}
\end{center}
\caption{Two-slit interferometer.}
\end{figure}
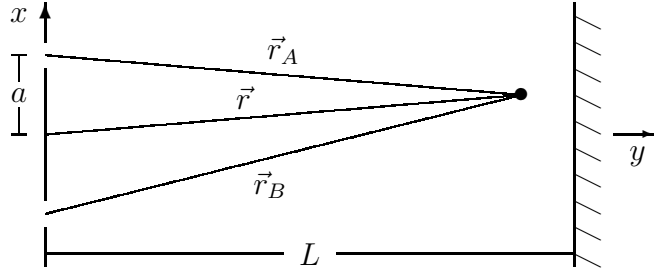

Since the particles are taken to be bosons, the
wave function at a specific time is given 
(up to a constant factor) as in
Eq.~(8) of Ref.~\cite{ghosea}:
\begin{equation}
\Psi (x_1, y_1; x_2, y_2) =  
\psi _A (x_1, y_1) \psi _B (x_2, y_2) +
\psi _A (x_2, y_2) \psi _B (x_1, y_1) .
\label{eq3a}
\end{equation}
The one-particle wave functions, however, should
be represented by spherical, rather than plane
waves.  This corresponds to the fact that they
originate from the slits.
Referring to Fig.~1, we have
\begin{equation}
r_A = |\vec{r} - a \hat{x}|
= \sqrt{y^2 + (x-a)^2} \approx
y + \frac{1}{2L} (x-a)^2 ,
\label{eq3b}
\end{equation}
where the fact that $y \approx L$ has been
used, and higher order terms have been
neglected.  Similarly,
\begin{equation}
r_B  \approx y + \frac{1}{2L} (x+a)^2 .
\label{eq3c}
\end{equation}

For spherical waves we have
\begin{align}
\psi _A (\vec{r} \,) &= \frac{1}{r _A}
\exp (i k r_A) \approx \frac{1}{L}
\frac{1}{1 + \frac{y-L}{L}}
\exp \left\{ i k \left[ y + \frac{1}{2L}
(x-a)^2 \right] \right\} , \label{eq3d} \\
\psi _B (\vec{r} \,) &= \frac{1}{r _B}
\exp (i k r_B) \approx \frac{1}{L}
\frac{1}{1 + \frac{y-L}{L}}
\exp \left\{ i k \left[ y + \frac{1}{2L}
(x+a)^2 \right] \right\} .
\label{eq3e}
\end{align}
Substituting Eqs.~(\ref{eq3d}) and~(\ref{eq3e})
into (\ref{eq3a}), we get
\begin{align}
\Psi &= \frac{1}{L^2} \left(1 + \frac{y_1-L}{L} \right) ^{-1}
\left( 1 + \frac{y_2-L}{L} \right) ^{-1} \exp \left\{
i k (y_1 + y _2) \right\} \notag \\
& \qquad \cdot \left[ 
\exp \left\{ \frac{ik}{2L} \left[ (x_1 - a)^2 +
(x_2 + a)^2 \right] \right\} \right. \notag \\
& \qquad \quad \left. \mbox{} +
\exp \left\{ \frac{ik}{2L} \left[ (x_1 + a)^2 +
(x_2 - a)^2 \right] \right\} \right] . 
\label{eq3f}
\end{align}
From (\ref{eq3f}) we easily find that
\begin{equation}
\frac{dx_1}{dt} + \frac{dx_2}{dt} =
\frac{\hbar}{m} \mbox{Im} \left\{ \frac{1}{\Psi}
\left( \frac{\partial \Psi}{\partial x_1} + 
\frac{\partial \Psi}{\partial x_2} \right) \right\} =
\frac{\hbar k}{m L} (x_1 + x _2) .
\label{eq3g}
\end{equation}
Contrary to Ghose's claim, $\dot{x} _1 + \dot{x} _2$
does not vanish.  Therefore, his argument does not
go through.
\end{document}